\documentclass[traditabstract]{aa}  
\usepackage{graphicx}
\usepackage{natbib}
\begin{document}

   \title{The GAPS Programme with HARPS-N@TNG \\ 
          II: No giant planets around the metal-poor star HIP 11952
          \thanks{Based on observations made with the Italian Telescopio 
                  Nazionale Galileo (TNG) operated on the island of La Palma 
                  by the Fundacion Galileo Galilei of the INAF 
                  at the Spanish 
                  Observatorio del Roque de los Muchachos of the IAC
                  in the frame of the program 
                  Global Architecture of Planetary Systems (GAPS).
                  Based on observations collected at the La Silla Observatory, ESO (Chile):
                  Program 185.D-0056.}}

\titlerunning{GAPS II: No giant planets around HIP 11952}

   \author{S. Desidera 
           \inst{1},
           A. Sozzetti
           \inst{2},
           A.S. Bonomo
           \inst{2},
           R. Gratton
           \inst{1},
           E. Poretti
           \inst{3},
           R. Claudi
           \inst{1},
           D.W. Latham
           \inst{4}, \\
           L. Affer
           \inst{5}, 
           R. Cosentino
           \inst{6,7},
           M. Damasso
           \inst{8,9,2},
           M. Esposito
           \inst{10,11},
           P. Giacobbe
           \inst{12,2},
           L. Malavolta
           \inst{8,13},\\
           V. Nascimbeni
           \inst{8,1}, 
           G. Piotto
           \inst{8,1},
           M. Rainer
           \inst{3},
           M. Scardia
           \inst{3},
           V.S. Schmid
           \inst{14},
           A.F. Lanza
           \inst{6},
           G. Micela
           \inst{5}, \\
           I. Pagano
           \inst{6},
           L. Bedin
           \inst{1},
           K. Biazzo
           \inst{15},
           F. Borsa
           \inst{3,16},
           E. Carolo
           \inst{1},
           E. Covino
           \inst{15},
           F. Faedi
           \inst{17},
           G. H\'ebrard
           \inst{18},\\ 
           C. Lovis
           \inst{13},
           A. Maggio
           \inst{5},
           L. Mancini
           \inst{19},
           F. Marzari
           \inst{20,1},
           S. Messina
           \inst{6}, 
           E. Molinari
           \inst{7,21},
           U. Munari
           \inst{1}, \\
           F. Pepe
           \inst{13},
           N. Santos
           \inst{22,23},
           G. Scandariato
           \inst{6},
           E. Shkolnik
           \inst{24},
           J. Southworth
           \inst{25}
           }

   \authorrunning{S. Desidera et al.}

   \offprints{S. Desidera} 

   \institute{INAF -- Osservatorio Astronomico di Padova,  
              Vicolo dell'Osservatorio 5, I-35122, Padova, Italy
             \and 
             INAF -- Osservatorio Astrofisico di Torino, 
             Via Osservatorio 20, I-10025, Pino Torinese, Italy
             \and
             INAF -- Osservatorio Astronomico di Brera, Via E. Bianchi 46, I-23807 Merate (LC), Italy,
             \and
          Harvard-Smithsonian Center for Astrophysics, 60 Garden Street, Cambridge, MA 02138
             \and
             INAF -- Osservatorio Astronomico di Palermo, Piazza del Parlamento, Italy 1, I-90134, Palermo, Italy
             \and
            INAF -- Osservatorio Astrofisico di Catania, Via S.Sofia 78, I-9512, Catania, Italy
             \and
             Fundaci\'on Galileo Galilei - INAF,
             Rambla Jos\'e Ana Fernández P\'erez, 7
             38712 Bre\~na Baja, TF - Spain
             \and
            Dip. di Fisica e Astronomia Galileo Galilei -- Universit\`a di Padova, Vicolo
             dell'Osservatorio 2, I-35122, Padova, Italy 
            \and
            Osservatorio Astronomico della Regione Autonoma Valle d'Aosta, 
            Fraz. Lignan 39, 11020, Nus (Aosta), Italy
             \and
             Instituto de Astrofisica de Canarias, C/Via Lactea S/N, E-38200 La Laguna, Tenerife, Spain
             \and
            Departamento de Astrofisica, Universidad de La Laguna,  E-38205 La  
            Laguna, Tenerife, Spain
             \and
            Dipartimento di Fisica, Universit\`a  di Trieste, Via Tiepolo 11, I-34143 Trieste, Italy
             \and
           Observatoire Astronomique de l'Universit\'e de Geneve, 51 ch. des Maillettes - 
           Sauverny, 1290, Versoix, Switzerland
             \and
            Instituut voor Sterrenkunde, K.U.\ Leuven, Celestijnenlaan 200D, 3001 Leuven, Belgium
             \and
            INAF -- Osservatorio Astronomico di Capodimonte, Salita Moiariello 16, I-80131, Napoli, Italy
             \and
           Dipartimento di Scienza e Alta Tecnologia, Universit\`a dell'Insubria, Via Valleggio 11, 22100 Como, Italy
              \and  
          Department of Physics, University of Warwick, Gibbet Hill Road, Coventry, CV4 7AL, UK
             \and
         Institut d'Astrophysique de Paris, UMR7095 CNRS, Univ. 
          P. \& M. Curie, 98bis bd.\ Arago, F-75014 Paris, France
             \and
         Max-Planck-Institut f\"ur Astronomie, K\"onigstuhl 17, D-69117, Heidelberg, Germany
            \and
            Dipartimento di Fisica e Astronomia Galileo Galilei -- Universit\`a di Padova, Via Marzolo 8, Padova, Italy 
            \and
           INAF - IASF Milano, via Bassini 15, I-20133 Milano, Italy
            \and
      Centro de Astrof{\'\i}sica, Universidade do Porto, Rua das Estrelas, 
      4150-762 Porto, Portugal
           \and
     Departamento de F{\'\i}sica e Astronomia, Faculdade de Ci\^encias,  Universidade do Porto, Portugal
          \and
       Lowell Observatory, 1400 W. Mars Hill Road, Flagstaff, AZ 86001, USA
          \and
        Astrophysics Group, Keele University, Staffordshire, ST5 5BG, UK
              }

\date{Received  / Accepted }

\abstract{In the context of the program 
Global Architecture of Planetary Systems (GAPS),
we have performed radial velocity monitoring of the metal-poor star 
HIP 11952 on 35 nights over about 150 days using the newly installed high resolution spectrograph 
HARPS-N at the TNG and HARPS at ESO 3.6m telescope.
The radial velocities show a scatter of 7 m s$^{-1}$, compatible with the measurement 
errors for such a moderately warm metal-poor star (T$_\mathrm{eff} = 6040\pm120$~K; [Fe/H] $=-1.9\pm0.1$). 
We then exclude the presence of the two giant planets
with periods of $6.95\pm0.01$ d and $290.0\pm16.2$ d and radial velocity semi-amplitudes of 
$100.3\pm19.4$ m s$^{-1}$ and $105.2\pm14.7$ m s$^{-1}$, 
respectively, which had recently been announced. 
This result is important considering that HIP 11952 was thought to be the most metal-poor star
hosting a planetary system with giant planets, thus challenging some models of planet formation.}


   \maketitle

\section{Introduction}
\label{s:intro}

The identification of planets around stars with abundances of heavy element significantly lower than the Sun 
represents a relevant test for models of the formation of planetary systems.
On the one hand, assuming stellar metallicity ([Fe/H]) to be a natural proxy for the actual heavy metal content of the primordial circumstellar disk, 
the core-accretion scenario for giant planet formation predicts that the planet frequency $f_{\rm p}$ should increase with higher 
[Fe/H]~\citep{2004ApJ...616..567I,2009A&A...501.1139M}, 
as the increased surface density of solids facilitates the growth of embryonic cores, thus greatly enhancing the formation of gas giant planets 
around more metal-rich stars. Indeed, there exist theoretical expectations for threshold values of [Fe/H] below which giant planets 
cannot form, [Fe/H] $\simeq -0.5$ \citep{2012ApJ...751...81J,2012A&A...541A..97M}. 
On the other hand, gas giant planet formation by gravitational instability is less sensitive
to the disk metal content, thus $f_{\rm p}$ is expected to be 
independent of [Fe/H]~\citep{2002ApJ...567L.149B}. 

The observational evidence of a strong dependence of $f_{\rm p}$ on [Fe/H] for giant planets \citep{2004A&A...415.1153S,2005ApJ...622.1102F} is 
commonly considered as supporting the core-accretion mechanism \citep{2009A&A...501.1161M}.
No significant trends of $f_p$\ with [Fe/H] are observed for low-mass planets (Neptunes and
super-Earths) in radial-velocity (RV) surveys, for the metal abundance range $-0.5\leq$[Fe/H]$\leq +0.1$ 
\citep{2011arXiv1109.2497M,2011A&A...533A.141S}. This result was recently confirmed by statistical 
analyses of {\it Kepler} transiting planet candidates \citep{2012Natur.486..375B}.
Moreover, low-mass planets seem to be rare around super-metal-rich stars \citep{2012arXiv1207.1012J}.

The $f_{\rm p}$--[Fe/H] relation for giant planets is firmly established on a solid statistical basis for [Fe/H] $> 0.0$. 
In this regime, a simple power-law fit with $\alpha\times10^{\beta\mathrm{[Fe/H]}}$, and $\beta$ in the range $1.5$--$2.0$, 
well represents the observed trend \citep[e.g][]{2005ApJ...622.1102F,2009ApJ...697..544S,2010PASP..122..905J}. 
Stars with [Fe/H] $\sim+0.3$ appear $4$--$5$ times more likely to host a giant planet than solar-metallicity dwarfs. At 
[Fe/H] $\simeq0.0$, $f_p$ is in the neighbourhood of $3$--$5$\%. 
The situation is however less clear for [Fe/H] $< 0.0$, and recent studies~\citep{2012A&A...543A..45M} indicate the possibility 
that a power law might not describe correctly the relation in the low-metallicity regime.

The uncertainties still apparent in statistical studies stem primarily from the relatively limited sample sizes of 
metal-poor stars in large RV surveys. Attempts at mitigating this limitation have been made in the past, with experiments 
focusing on RV searches for giant planets around about 250 metal-poor stars carried out by~\citet{2006ApJ...649..428S,2009ApJ...697..544S} 
with Keck/HIRES and~\citet{2011A&A...526A.112S} using HARPS on the ESO 3.6m telescope. The outcome of the first survey 
was a null result, 
while the HARPS survey yielded three detections at the metal-rich end ([Fe/H] $\sim -0.5$) of 
the sample~\citep{2007A&A...474..647S,2010A&A...512A..47S}.
Another project started in June 2009 around a sample of 96 metal-poor stars with the FEROS spectrograph. 
Detections have been reported from this survey of a short-period ($P=16$ d) giant planet around 
the metal-poor horizontal branch star HIP 13044~\citep{2010Sci...330.1642S}
and of a two-planet system orbiting the F dwarf HIP 11952~\citep[][hereafter S12]{2012A&A...540A.141S}.

The HIP 11952 system is of particular interest. The primary is a high-proper-motion, nearby ($d=112$ pc), 
relatively bright ($V=9.85$) early F-type dwarf (possibly a subgiant), with [Fe/H] $=-1.9\pm0.1$ (S12). 
The two planets reported by S12 have $P=6.95\pm0.01$ d and 
and minimum masses $m_2\sin i = 0.78\pm0.16$ M$_\mathrm{J}$ and 
$2.93\pm0.42$ M$_\mathrm{J}$, respectively. 
The HIP 11952 system, with a metallicity ten times lower than the second-lowest metallicity giant-planet hosting dwarf, 
poses a severe challenge to the core-accretion model~\citep{2012ApJ...751...81J,2012A&A...541A..97M}.  

HIP 11952 was included in our program focusing on known planetary systems within the framework
of the long-term project {\it Global Architecture of Planetary Systems} (GAPS) recently started
in open time using HARPS-N at the Telescopio Nazionale Galileo (TNG). A description of the program is provided 
in Covino et al.\ (A\&A to be submitted).
We aimed at a) improving the quality of the orbital solutions reported by S12, and b) looking for evidence of 
lower-mass companions. In this Letter, based on RV measurements with typical internal errors $5$--$10$ times 
lower than the originally published FEROS values, we report a non-detection of both giant planets announced by S12. 

\section{HARPS-N}
\label{s:harpsn}

HARPS-N is an \'echelle spectrograph covering the visible wavelength range between 383 and 693 nm~\citep{2012SPIE.8446E..1VC}. 
It is a near-twin of the HARPS instrument mounted at the ESO 3.6-m telescope in La Silla \citep{2003Msngr.114...20M}. 
It was installed at the TNG in spring 2012. After instrument commissioning
in late spring and summer 2012, it was offered for open time programs starting in August 2012.

The instrument is located in a thermally-controlled room, within a vacuum-controlled enclosure
to ensure the required stability, and is fed by two fibres at the Nasmyth B focus of the TNG. The second 
fibre can be used for simultaneous calibration (currently with a Th-Ar hollow-cathode lamp) or 
for monitoring of the sky depending on the science goal and target brightness.
Both fibres have an aperture on the sky of 1 arcsec.
The spectra are recorded on an E2V 4k4 CCD 231 with a $15~{\rm \mu m}$ pixel
size. The resulting sampling is about 3.3 pixels (FWHM) and the spectral resolution is about 115,000.
Early observing tests yielded an instrument total efficiency of $\varepsilon = 7$\% at 550 nm, 
including losses due to the Earth's atmosphere and the telescope mirrors. 

A failure of the red side of the CCD in late Sept 2012 caused the observations between Sept 29th and Oct 25th 2012 
to be performed using the blue side only. A new CCD was installed at the beginning of November 2012, and observations after
this date were performed with the full spectral range.
The possible impact of the observations taken with only half of the spectral range is discussed in \S~\ref{s:rv}.

\section{Observations and data reduction}
\label{s:obsred}

HIP 11952 was observed with HARPS-N on 25 nights from 2012 Aug 7 to 2013 Jan 6.
The Th-Ar simultaneous calibration was used in all observations.
The drift correction with respect to the reference calibration was almost always below 1 m s$^{-1}$.
The integration time of 900 s adopted for all the observations led to a
typical S/N ratio of $\sim70$--$80$ per pixel on the extracted spectrum at 460 nm.
Additional observations on 14 nights in two runs from 2012 Dec 11 to 2013 Jan 5 were also gathered 
at the ESO 3.6m telescope using HARPS. An integration time of 1200 s was adopted. 

RV measurements, and their internal errors, were obtained using the online pipelines of HARPS and HARPS-N, which are 
based on the  
weighted cross-correlation function (CCF) method \citep{1996A&AS..119..373B,2002A&A...388..632P}. 
The G2 mask was adopted (the earliest-type one available). The median of the internal errors is $\sigma_\mathrm{RV} = 5.8$ m s$^{-1}$ 
for HARPS-N (full CCDs in use) and $\sigma_\mathrm{RV} = 5.6$ m s$^{-1}$ for HARPS. The relatively large RV errors are due to the 
paucity of spectral lines of such a moderately warm, metal-poor star. 
The HARPS-N  spectra between JD 2456199 and 2456228 were acquired with only the
spectral orders falling on the blue side of the CCD.
Besides the increase of the RV uncertainties due to the lower number of lines (median $\sigma_\mathrm{RV} = 8.3$ m s$^{-1}$),
using only half of the spectral range also introduced a systematic RV zero-point shift, with possible additional 
shifts due to the slight change in dispersion and spectral resolution required for the 
optimisation of the \'echelle grating angle.
We estimated such a shift by deriving the RVs of full-chip spectra including only the spectral
orders on the blue side of the CCD, i.e.\ with $\lambda < 534.75$~nm.
The average offset is $7.4\pm0.9$ m s$^{-1}$, the half-chip RVs being higher. 
Such a shift is consistent with RV differences observed for other stars in the GAPS program using the same mask.
Additional offsets should also be present between the RVs obtained with HARPS and HARPS-N, as they cover slightly
different spectral ranges, and possibly for the RVs taken with the new HARPS-N CCD since focus adjustments 
were made and chip properties may be different. However, from the available data, these offsets appear 
below the measurement uncertainties. 

\section{Analysis}
\label{s:rv}

\begin{figure}
   \includegraphics[width=7.8cm]{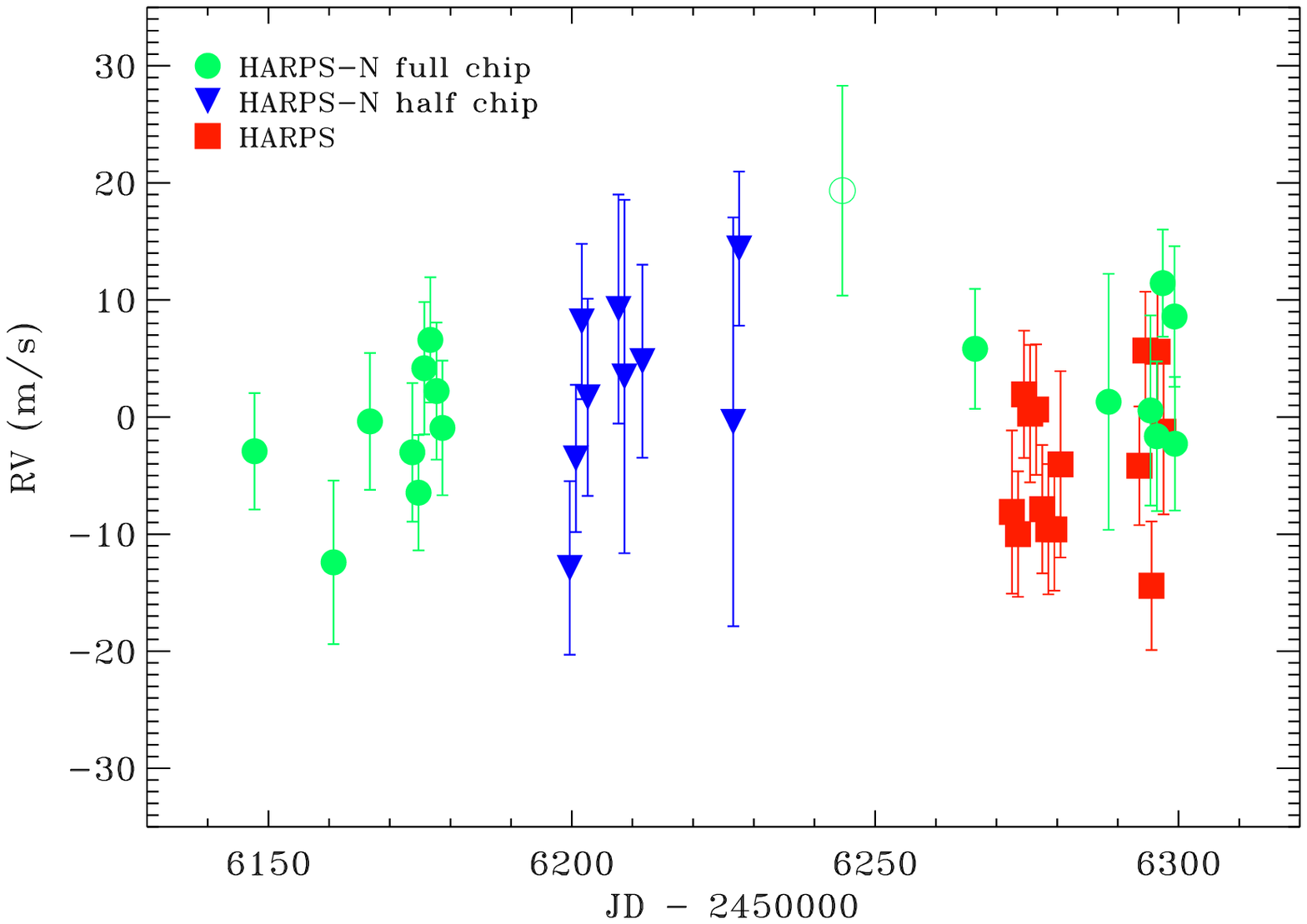}
   \includegraphics[width=7.8cm]{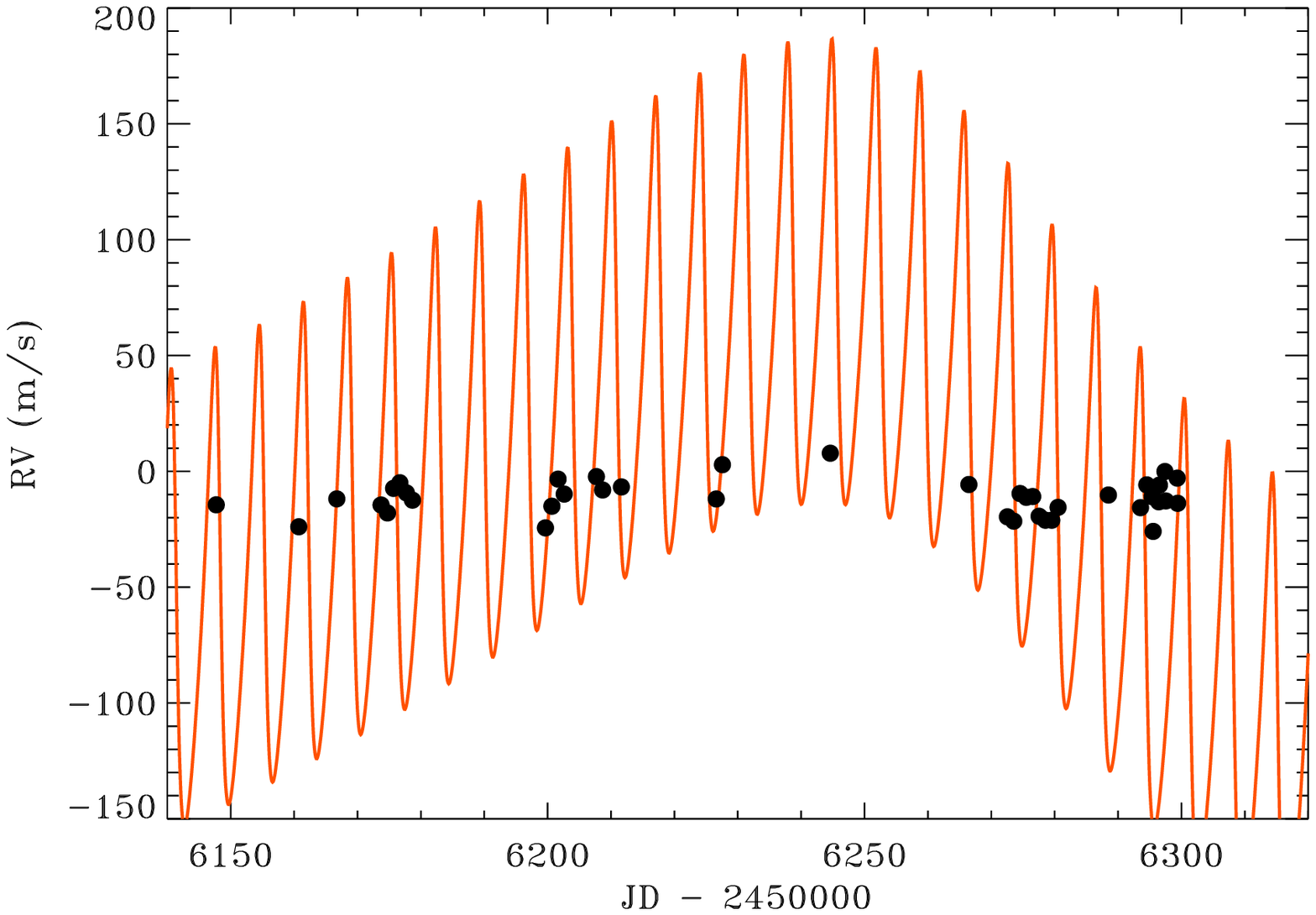}
      \caption{Relative RVs of \object{HIP 11952} obtained with HARPS-N and HARPS. 
               In the upper panel, green circles represent full-chip HARPS-N data, blue triangles blue-chip
               HARPS-N data and red squares HARPS data. An offset of $-7.4$ m/s was applied to the blue-chip data.
               The open green circle is not included in the analysis. 
               In the lower panel the predicted
               RV signature of the two planets by S12 is overplotted as a solid line.}        
         \label{f:rv}
   \end{figure}

The relative RV time series is shown in Fig.~\ref{f:rv}. We report in Table~\ref{t:rv} the full dataset. 
We included the correction of $+7.4$ m s$^{-1}$ for the half-chip RVs. We anticipate that, whilst this 
has an obvious impact when attempting to search for RV variations with an amplitude close to the measurement 
uncertainties or to investigate the existence of a long-term trend, the main result of the paper is entirely unaffected 
by this procedure. The RV measurement obtained on 2012 Nov 12 (the blue open circle at JD 2456244.5992 in the upper panel of Fig~.1), 
immediately after operations were resumed with the 
new HARPS-N CCD in place, is not included in the analysis as it has discrepant values of CCF contrast and activity index, 
pointing to a temporary calibration problem with the new setup. 

\begin{figure}
   \includegraphics[width=7.8cm]{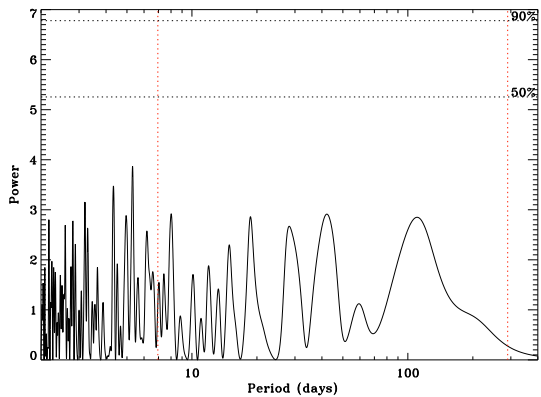}
      \caption{Lomb-Scargle periodogram of RVs of \object{HIP 11952} 
               obtained with HARPS-N and HARPS. 
               Confidence levels from the bootstrap simulations are shown.
               The vertical dotted lines mark the periodicities identified
               by S12 based on FEROS RVs.}        
         \label{f:scargle}
   \end{figure}

The offset-corrected scatter (6.9 m s$^{-1}$) is consistent with the internal errors.
Considering our sampling, RV variations with periods and semi-amplitudes ($\sim100$ m s$^{-1}$) 
close to those reported in~\citet{2012A&A...540A.141S} are clearly ruled out.
A Lomb-Scargle periodogram of the RVs does not indicate any significant periodicities
in the range $2$--$400$ days (Fig.~\ref{f:scargle}). The confidence levels in Fig.~\ref{f:scargle} 
were obtained by random permutations as in~\citet{2011A&A...533A..90D}.

Furthermore, there are neither significant periodicities nor correlations of the RVs with 
the CCF FWHM, the CCF contrast, and the bisector velocity span, as obtained by the instruments' pipelines,
as well as with the Ca II H\&K activity indicator, measured as in~\citet{2006A&A...454..553D}.
S12 also suggest that HIP 11952 could be a pulsating variable. This possibility
would be of great relevance, since in this case the RV measurements should primarily show 
the effect of the pulsation. The power spectra of the {\sc Hipparcos} and ASAS
\citep[All-Sky Automated Survey;][]{2002AcA....52..397P} photometric 
data are very noisy due to their poor spectral windows, but no significant periodic 
signal is detected. 

Our RVs do not show significant long-term trends. 
To further constrain the existence of a stellar companion on a long-period orbit, we considered
the 14 RVs obtained by \citet{2002AJ....124.1144L} with the CfA Digital Speedometers (DS) 
between 1984 and 1998. Typical errors are 0.7 km s$^{-1}$ and no obvious
trend in the RVs over that time span is apparent. Furthermore, Latham (private communication)
obtained five new RVs with the Tillinghast Reflector Echelle Spectrograph (TRES)
at the FLWO observatory over a span of eight nights
during December 2012 with typical errors of 0.1 km s$^{-1}$. The average
velocity for these RVs, when shifted to the velocity zero
point of the CfA DS (determined from extensive
observations of standard stars), is $23.90 \pm 0.05$ km s$^{-1}$ (uncertainty
of the mean), compared to $23.64 \pm 0.25$ km s$^{-1}$ for the 14 old RVs from
the CfA DS. The probability that the star has a
constant velocity, given the internal errors of the 19 observations,
is P($\chi^2$) $=$ 0.47. Thus there is no evidence for a secular drift in
the velocities of HIP 11952 over a span of 29 years.

\begin{figure}
    \includegraphics[width=6.0cm,angle=90]{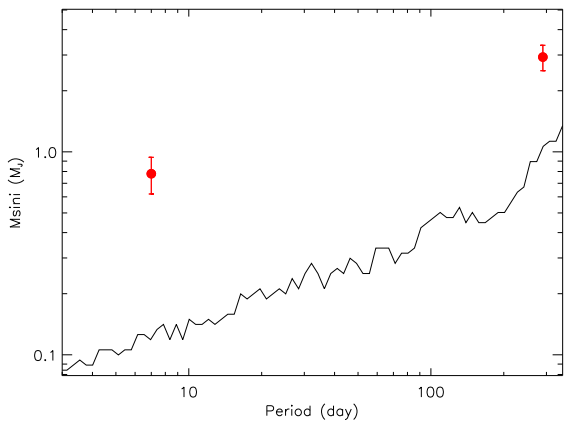}
       \caption{Upper limits on the minimum mass of possible planetary
companions around HIP 11952 with 99\% confidence level from HARPS-N and HARPS
data. Filled circles show the mass and period of the two planets claimed by
 S12 that are clearly incompatible with our data.}
          \label{f:limits}
    \end{figure}

Following~\citet{2009ApJ...697..544S}, upper limits on the minimum mass of possible planetary 
companions with $3\leq P\leq 350$ days and eccentricity $e<0.6$ were derived from our data with 99\%
confidence level, based on the F-test and $\chi^2$ statistics (Fig.~\ref{f:limits}).
They show that both the inner and outer planets claimed by S12 are ruled out by our observations, at 
the 6-$\sigma$ and 4-$\sigma$ level, respectively.
Actually, this test excludes at the given level of 
confidence {\it any} planet with masses and periods as given by S12, whatever their orbital parameters. 
Planets with masses, periods and other orbital parameters as given by S12 
are excluded at higher levels of confidence by our data. 

\section{Discussion and conclusions}
\label{s:discussion}

Based on high-precision RV measurements carried out with HARPS-N and HARPS we found no evidence of 
the two giant planets with $P=6.95$ d and $P=290$ d, and projected masses $0.78\pm0.16$ M$_\mathrm{J}$ and $2.93\pm0.42$ M$_{J}$, 
respectively, announced by S12 orbiting the metal-poor star HIP 11952.
Saturn-mass and Jupiter-mass planets with periods shorter than 30 and 250 days, respectively, are also excluded. 
Our observations are within the measurement errors, that are about $5$--$10$ times lower than those of the FEROS
observations, thanks to the optimised performance of HARPS-N in delivering RVs and
the larger telescope aperture. 
This case shows that care should be taken in interpreting RV variations as indicating orbital motion,
as stellar and instrumental effects could impact the measurements, particularly when inferred RV amplitudes
are only slightly larger than the measurement accuracy.

Our result has important consequences, as it clears the observational sample of a system
that constituted a severe challenge to the core-accretion model of giant planet formation. A giant planet system around 
HIP 11952 was indeed very hard to explain within the context of the most recent calculations based on this 
mechanism~\citep{2012ApJ...751...81J,2012A&A...541A..97M}.

On the one hand, giant planets formed by core accretion are expected to be very rare around low-metallicity 
dwarfs, and the observational evidence presented here directly supports this view. On the other hand, 
low-mass planets can theoretically form around metal-poor stars, so it is highly desirable to obtain 
statistically useful observational inferences 
on the actual 
value of $f_p$ for Super-Earths and Neptunes across orders of magnitude in the host stars' metal content. 
New programs to search for low-mass planets around metal-poor stars 
(such as the on going ESO large program 190.C-0027 on HARPS and our dedicated GAPS program
on HARPS-N)
will then allow for a significant improvements in our knowledge of the relative roles of 
competing planet formation processes.

\begin{acknowledgements}

We thank the TNG staff for help in the preparation of the observations and 
R.\ Smareglia and collaborators for the kind assistance with the 
data retrieval from TNG archive.
This work was partially funded by PRIN-INAF 2010.
NCS acknowledges the support by the ERC/EC 
under the FP7 through Starting Grant agreement n.~239953,
as well as the support from Funda\c{c}\~ao para a Ci\^encia e a 
Tecnologia (FCT) through prog.~Ci\^encia\,2007 funded by FCT/MCTES 
(Portugal) and POPH/FSE (EC), and in the form of grants ref.~PTDC/CTE-AST/098528/2008 and PTDC/CTE-AST/120251/2010.

\end{acknowledgements}

\bibliography{hip11952}

\begin{thebibliography}{29}
\expandafter\ifx\csname natexlab\endcsname\relax\def\natexlab#1{#1}\fi

\bibitem[{{Baranne} {et~al.}(1996){Baranne}, {Queloz}, {Mayor}, {Adrianzyk},
  {Knispel}, {Kohler}, {Lacroix}, {Meunier}, {Rimbaud}, \&
  {Vin}}]{1996A&AS..119..373B}
{Baranne}, A., {Queloz}, D., {Mayor}, M., {et~al.} 1996, \aaps, 119, 373

\bibitem[{{Boss}(2002)}]{2002ApJ...567L.149B}
{Boss}, A.~P. 2002, \apjl, 567, L149

\bibitem[{{Buchhave} {et~al.}(2012){Buchhave}, {Latham}, {Johansen},
  {Bizzarro}, {Torres}, {Rowe}, {Batalha}, {Borucki}, {Brugamyer}, {Caldwell},
  {Bryson}, {Ciardi}, {Cochran}, {Endl}, {Esquerdo}, {Ford}, {Geary},
  {Gilliland}, {Hansen}, {Isaacson}, {Laird}, {Lucas}, {Marcy}, {Morse},
  {Robertson}, {Shporer}, {Stefanik}, {Still}, \&
  {Quinn}}]{2012Natur.486..375B}
{Buchhave}, L.~A., {Latham}, D.~W., {Johansen}, A., {et~al.} 2012, \nat, 486,
  375

\bibitem[{{Cosentino} {et~al.}(2012){Cosentino}, {Lovis}, {Pepe}, {Collier
  Cameron}, {Latham}, {Molinari}, {Udry}, {Bezawada}, {Black}, {Born},
  {Buchschacher}, {Charbonneau}, {Figueira}, {Fleury}, {Galli}, {Gallie},
  {Gao}, {Ghedina}, {Gonzalez}, {Gonzalez}, {Guerra}, {Henry}, {Horne},
  {Hughes}, {Kelly}, {Lodi}, {Lunney}, {Maire}, {Mayor}, {Micela}, {Ordway},
  {Peacock}, {Phillips}, {Piotto}, {Pollacco}, {Queloz}, {Rice}, {Riverol},
  {Riverol}, {San Juan}, {Sasselov}, {Segransan}, {Sozzetti}, {Sosnowska},
  {Stobie}, {Szentgyorgyi}, {Vick}, \& {Weber}}]{2012SPIE.8446E..1VC}
{Cosentino}, R., {Lovis}, C., {Pepe}, F., {et~al.} 2012, in Society of
  Photo-Optical Instrumentation Engineers (SPIE) Conference Series, Vol. 8446,
  Society of Photo-Optical Instrumentation Engineers (SPIE) Conference Series

\bibitem[{{Desidera} {et~al.}(2011){Desidera}, {Carolo}, {Gratton}, {Martinez
  Fiorenzano}, {Endl}, {Mesa}, {Barbieri}, {Bonavita}, {Cecconi}, {Claudi},
  {Cosentino}, {Marzari}, \& {Scuderi}}]{2011A&A...533A..90D}
{Desidera}, S., {Carolo}, E., {Gratton}, R., {et~al.} 2011, \aap, 533, A90

\bibitem[{{Desidera} {et~al.}(2006){Desidera}, {Gratton}, {Lucatello},
  {Claudi}, \& {Dall}}]{2006A&A...454..553D}
{Desidera}, S., {Gratton}, R.~G., {Lucatello}, S., {Claudi}, R.~U., \& {Dall},
  T.~H. 2006, \aap, 454, 553

\bibitem[{{Fischer} \& {Valenti}(2005)}]{2005ApJ...622.1102F}
{Fischer}, D.~A. \& {Valenti}, J. 2005, \apj, 622, 1102

\bibitem[{{Ida} \& {Lin}(2004)}]{2004ApJ...616..567I}
{Ida}, S. \& {Lin}, D.~N.~C. 2004, \apj, 616, 567

\bibitem[{{Jenkins} {et~al.}(2012){Jenkins}, {Jones}, {Tuomi}, {Murgas},
  {Hoyer}, {Jones}, {Barnes}, {Pavlenko}, {Ivanyuk}, {Rojo}, {Jordan},
  {Day-Jones}, {Ruiz}, \& {Pinfield}}]{2012arXiv1207.1012J}
{Jenkins}, J.~S., {Jones}, H.~R.~A., {Tuomi}, M., {et~al.} 2012, ArXiv e-prints

\bibitem[{{Johnson} {et~al.}(2010){Johnson}, {Aller}, {Howard}, \&
  {Crepp}}]{2010PASP..122..905J}
{Johnson}, J.~A., {Aller}, K.~M., {Howard}, A.~W., \& {Crepp}, J.~R. 2010,
  \pasp, 122, 905

\bibitem[{{Johnson} \& {Li}(2012)}]{2012ApJ...751...81J}
{Johnson}, J.~L. \& {Li}, H. 2012, \apj, 751, 81

\bibitem[{{Latham} {et~al.}(2002){Latham}, {Stefanik}, {Torres}, {Davis},
  {Mazeh}, {Carney}, {Laird}, \& {Morse}}]{2002AJ....124.1144L}
{Latham}, D.~W., {Stefanik}, R.~P., {Torres}, G., {et~al.} 2002, \aj, 124, 1144

\bibitem[{{Mayor} {et~al.}(2011){Mayor}, {Marmier}, {Lovis}, {Udry},
  {S{\'e}gransan}, {Pepe}, {Benz}, {Bertaux}, {Bouchy}, {Dumusque}, {Lo Curto},
  {Mordasini}, {Queloz}, \& {Santos}}]{2011arXiv1109.2497M}
{Mayor}, M., {Marmier}, M., {Lovis}, C., {et~al.} 2011, ArXiv e-prints

\bibitem[{{Mayor} {et~al.}(2003){Mayor}, {Pepe}, {Queloz}, {Bouchy},
  {Rupprecht}, {Lo Curto}, {Avila}, {Benz}, {Bertaux}, {Bonfils}, {Dall},
  {Dekker}, {Delabre}, {Eckert}, {Fleury}, {Gilliotte}, {Gojak}, {Guzman},
  {Kohler}, {Lizon}, {Longinotti}, {Lovis}, {Megevand}, {Pasquini}, {Reyes},
  {Sivan}, {Sosnowska}, {Soto}, {Udry}, {van Kesteren}, {Weber}, \&
  {Weilenmann}}]{2003Msngr.114...20M}
{Mayor}, M., {Pepe}, F., {Queloz}, D., {et~al.} 2003, The Messenger, 114, 20

\bibitem[{{Mordasini} {et~al.}(2009{\natexlab{a}}){Mordasini}, {Alibert}, \&
  {Benz}}]{2009A&A...501.1139M}
{Mordasini}, C., {Alibert}, Y., \& {Benz}, W. 2009{\natexlab{a}}, \aap, 501,
  1139

\bibitem[{{Mordasini} {et~al.}(2012){Mordasini}, {Alibert}, {Benz}, {Klahr}, \&
  {Henning}}]{2012A&A...541A..97M}
{Mordasini}, C., {Alibert}, Y., {Benz}, W., {Klahr}, H., \& {Henning}, T. 2012,
  \aap, 541, A97

\bibitem[{{Mordasini} {et~al.}(2009{\natexlab{b}}){Mordasini}, {Alibert},
  {Benz}, \& {Naef}}]{2009A&A...501.1161M}
{Mordasini}, C., {Alibert}, Y., {Benz}, W., \& {Naef}, D. 2009{\natexlab{b}},
  \aap, 501, 1161

\bibitem[{{Mortier} {et~al.}(2012){Mortier}, {Santos}, {Sozzetti}, {Mayor},
  {Latham}, {Bonfils}, \& {Udry}}]{2012A&A...543A..45M}
{Mortier}, A., {Santos}, N.~C., {Sozzetti}, A., {et~al.} 2012, \aap, 543, A45

\bibitem[{{Pepe} {et~al.}(2002){Pepe}, {Mayor}, {Galland}, {Naef}, {Queloz},
  {Santos}, {Udry}, \& {Burnet}}]{2002A&A...388..632P}
{Pepe}, F., {Mayor}, M., {Galland}, F., {et~al.} 2002, \aap, 388, 632

\bibitem[{{Pojmanski}(2002)}]{2002AcA....52..397P}
{Pojmanski}, G. 2002, \actaa, 52, 397

\bibitem[{{Santos} {et~al.}(2004){Santos}, {Israelian}, \&
  {Mayor}}]{2004A&A...415.1153S}
{Santos}, N.~C., {Israelian}, G., \& {Mayor}, M. 2004, \aap, 415, 1153

\bibitem[{{Santos} {et~al.}(2010){Santos}, {Mayor}, {Benz}, {Bouchy},
  {Figueira}, {Lo Curto}, {Lovis}, {Melo}, {Moutou}, {Naef}, {Pepe}, {Queloz},
  {Sousa}, \& {Udry}}]{2010A&A...512A..47S}
{Santos}, N.~C., {Mayor}, M., {Benz}, W., {et~al.} 2010, \aap, 512, A47

\bibitem[{{Santos} {et~al.}(2011){Santos}, {Mayor}, {Bonfils}, {Dumusque},
  {Bouchy}, {Figueira}, {Lovis}, {Melo}, {Pepe}, {Queloz}, {S{\'e}gransan},
  {Sousa}, \& {Udry}}]{2011A&A...526A.112S}
{Santos}, N.~C., {Mayor}, M., {Bonfils}, X., {et~al.} 2011, \aap, 526, A112

\bibitem[{{Santos} {et~al.}(2007){Santos}, {Mayor}, {Bouchy}, {Pepe}, {Queloz},
  \& {Udry}}]{2007A&A...474..647S}
{Santos}, N.~C., {Mayor}, M., {Bouchy}, F., {et~al.} 2007, \aap, 474, 647

\bibitem[{{Setiawan} {et~al.}(2010){Setiawan}, {Klement}, {Henning}, {Rix},
  {Rochau}, {Rodmann}, \& {Schulze-Hartung}}]{2010Sci...330.1642S}
{Setiawan}, J., {Klement}, R.~J., {Henning}, T., {et~al.} 2010, Science, 330,
  1642

\bibitem[{{Setiawan} {et~al.}(2012){Setiawan}, {Roccatagliata}, {Fedele},
  {Henning}, {Pasquali}, {Rodr{\'{\i}}guez-Ledesma}, {Caffau}, {Seemann}, \&
  {Klement}}]{2012A&A...540A.141S}
{Setiawan}, J., {Roccatagliata}, V., {Fedele}, D., {et~al.} 2012, \aap, 540,
  A141

\bibitem[{{Sousa} {et~al.}(2011){Sousa}, {Santos}, {Israelian}, {Mayor}, \&
  {Udry}}]{2011A&A...533A.141S}
{Sousa}, S.~G., {Santos}, N.~C., {Israelian}, G., {Mayor}, M., \& {Udry}, S.
  2011, \aap, 533, A141

\bibitem[{{Sozzetti} {et~al.}(2006){Sozzetti}, {Torres}, {Latham}, {Carney},
  {Stefanik}, {Boss}, {Laird}, \& {Korzennik}}]{2006ApJ...649..428S}
{Sozzetti}, A., {Torres}, G., {Latham}, D.~W., {et~al.} 2006, \apj, 649, 428

\bibitem[{{Sozzetti} {et~al.}(2009){Sozzetti}, {Torres}, {Latham}, {Stefanik},
  {Korzennik}, {Boss}, {Carney}, \& {Laird}}]{2009ApJ...697..544S}
{Sozzetti}, A., {Torres}, G., {Latham}, D.~W., {et~al.} 2009, \apj, 697, 544

\end{thebibliography}
\bibliographystyle{aa}

\begin{table}
   \caption[]{Radial velocities of HIP~11952. The last column indicates
              the instrument and CCD status: F is HARPS-N original full chip; H is HARPS-N half chip only (blue side); N is HARPS-N new chip; S is HARPS at ESO 3.6m.
              The tabulated RVs do not include any offset between the data obtained with different observational setups.}
     \label{t:rv}
       \centering
       \begin{tabular}{cccc}
         \hline
         \noalign{\smallskip}
         BJD -2450000  &  RV & error & Instrument/CCD \\
                      &  (km s$^{-1}$) & (km s$^{-1}$) &  \\
         \noalign{\smallskip}
         \hline
         \noalign{\smallskip}
 
 6147.72121  &   24.2530 &   0.0050 & F \\ 
 6160.74007  &   24.2435 &   0.0070 & F \\ 
 6166.74103  &   24.2556 &   0.0058 & F \\
 6173.72276  &   24.2529 &   0.0059 & F \\
 6174.73851  &   24.2495 &   0.0049 & F \\
 6175.69620  &   24.2601 &   0.0056 & F \\
 6176.68467  &   24.2625 &   0.0053 & F \\
 6177.71773  &   24.2582 &   0.0059 & F \\
 6178.68027  &   24.2550 &   0.0058 & F \\
 6199.65825  &   24.2505 &   0.0074 & H \\
 6200.67111  &   24.2598 &   0.0063 & H \\
 6201.65011  &   24.2715 &   0.0066 & H \\
 6202.61782  &   24.2650 &   0.0084 & H \\
 6207.70025  &   24.2726 &   0.0098 & H \\
 6208.68055  &   24.2668 &   0.0151 & H \\
 6211.64234  &   24.2681 &   0.0082 & H \\
 6226.61110  &   24.2629 &   0.0175 & H \\
 6227.58001  &   24.2777 &   0.0066 & H \\
 6244.59921  &   24.2753 &   0.0090 & N \\
 6266.46650  &   24.2618 &   0.0051 & N \\   
 6272.54203  &   24.2478 &   0.0070 & S \\  
 6273.54629  &   24.2459 &   0.0054 & S \\   
 6274.53914  &   24.2579 &   0.0054 & S \\   
 6275.53456  &   24.2562 &   0.0059 & S \\   
 6276.53597  &   24.2566 &   0.0056 & S \\   
 6277.54178  &   24.2481 &   0.0055 & S \\   
 6278.54057  &   24.2464 &   0.0056 & S \\   
 6279.53845  &   24.2464 &   0.0052 & S \\   
 6280.53859  &   24.2519 &   0.0080 & S \\   
 6288.48127  &   24.2573 &   0.0109 & N \\ 
 6293.54512  &   24.2518 &   0.0051 & S \\
 6294.53344  &   24.2616 &   0.0050 & S \\ 
 6294.86975  &   24.2565 &   0.0081 & N \\ 
 6295.53899  &   24.2415 &   0.0055 & S \\
 6295.91839  &   24.2543 &   0.0064 & N \\    
 6296.53505  &   24.2615 &   0.0062 & S \\
 6296.88798  &   24.2674 &   0.0046 & N \\  
 6297.53385  &   24.2547 &   0.0071 & S \\
 6299.33693  &   24.2645 &   0.0060 & N \\ 
 6299.40581  &   24.2537 &   0.0057 & N \\ 
         \noalign{\smallskip}
         \hline
      \end{tabular}

\end{table}

\end{document}